\documentclass[12pt]{myarticle}

\usepackage{txfonts}
\usepackage[colorlinks]{hyperref}

\newcommand{\n}{\noindent}
\newtheorem{Theorem}{Theorem}

\newtheorem{lemma}{Lemma}

\newtheorem{ex}{Example}

\begin{document}

\title{On eccentricity version of Laplacian energy of a graph}

\author{Nilanjan De\corref{cor1}}
\ead{de.nilanjan@rediffmail.com}

\address{Department of
Basic Sciences and Humanities (Mathematics),\\ Calcutta Institute of Engineering and Management, Kolkata, India.}
\cortext[cor1]{Corresponding Author.}

\begin{abstract}
The energy of a graph $G$ is equal to the sum of absolute values of the eigenvalues of the adjacency matrix of $G$, whereas the Laplacian energy of a graph $G$ is equal to the sum of the absolute value of the difference between the eigenvalues of the Laplacian matrix of $G$ and average degree of the vertices of $G$. Motivated by the work from Sharafdini et al. [R. Sharafdini, H. Panahbar, Vertex weighted Laplacian graph energy and other topological indices. \textit{J. Math. Nanosci.} 2016, 6, 49–-57.], in this paper we investigate the eccentricity version of Laplacian energy of a graph $G$.

\medskip
\noindent \textsl{MSC (2010):} Primary: 05C05.\\
\end{abstract}
\begin{keyword}
Eccentricity; Eigenvalue; Energy (of graph); Laplacian energy; Topological index.
\end{keyword}
\maketitle

\section{Introduction}

Let $G$ be a simple graph with $n$ vertices and $m$ edges. Let the vertex and edge sets of $G$ are denoted by $V(G)$ and $E(G)$ respectively. The degree of a vertex $v$, denoted by ${{d}_{G}}(v)$, is the number of vertices adjacent to $v$. For any two vertices $u,v\in V(G)$, the distance between $u$ and $v$ is denoted by ${{d}_{G}}(u,v)$ and is given by the number of edges in the shortest path connecting $u$ and $v$. Also we denote the sum of distances between $v\in V(G)$ and every other vertices in $G$ by $d(x|G)$, i.e., $d(x|G)=\sum\nolimits_{v\in V(G)}{{{d}_{G}}(x,v)}$. The eccentricity of a vertex $v$, denoted by ${{\varepsilon }_{G}}(v)$, is the largest distance from $v$ to any other vertex $u$ of $G$. The total eccentricity of a graph is denoted by $\zeta (G)$ and is equal to sum of eccentricities of all the vertices of the graph.
Let $A=[{{a}_{ij}}]$ be the adjacency matrix of G and let ${{\lambda }_{1}},{{\lambda }_{2}},...,{{\lambda }_{n}}$ are eigenvalues of $A$ which are the eigenvalues of the graph $G$. The energy of a graph is introduced by Ivan Gutman in 1978 \cite{gutm78} and defined as the sum of the absolute values of its eigenvalues and is denoted by E(G). Thus
\[E(G)=\sum\limits_{i=1}^{n}{|{{\lambda }_{i}}|}.\]
A large number of results on the graph energy have been reported, see for instance \cite{nikl07,bala04,gutm01,liu07}. Motivated by the success of the theory of graph energy, other different energy like quantities have been proposed and studied by different researcher. Let $D(G)=[{{d}_{ij}}]$ be the diagonal matrix associated with the graph $G$, where ${{d}_{ii}}={d_G}({{v}_{i}})$ and ${{d}_{ij}}=0$ if $i\ne j$. Define $L(G)=D(G)-A(G)$, where $L(G)$ is called the Laplacian matrix of $G$. Let ${{\mu }_{1}},{{\mu }_{2}},...,{{\mu }_{n}}$ be the Laplacian eigenvalues of $G$. Then the Laplacian energy of $G$ is defined as \cite{gutm06}
\[LE(G)=\sum\limits_{i=1}^{n}{|{{\mu }_{i}}-\frac{2m}{n}|}.\]
Various study on Laplacian energy of graphs were reported in the literature \cite{hoss10,das16,zhou08,alek08,merr94}. Analogues to Laplacian energy of a graph a different new type of graph energy were introduced and in this present study, inspired by the work in \cite{shar16}, we investigate the eccentricity version of Laplacian energy of a graph denoted by $L{{E}_{\varepsilon }}(G)$. In this case, we define the Laplacian eccentricity matrix as ${{L}_{\varepsilon }}(G)=\varepsilon (G)-A(G)$, where $\varepsilon (G)=[{{e}_{ij}}]$ is the $(n\times n)$ diagonal matrix of $G$ with ${{e}_{ii}}={\varepsilon_G} ({{v}_{i}})$ and ${{e}_{ij}}=0$ if $i\ne j$. Here, ${\varepsilon_G} ({{v}_{i}})$ is the eccentricity of the vertex ${{v}_{i}},i=1,2,...,n$.  Let $\mu _{1}^{'},\mu _{2}^{'},...,\mu _{n}^{'}$ be the eigenvalues of the matrix ${{L}_{\varepsilon }}(G)$. Then the eccentricity version of Laplacian energy of $G$ is defined as
\[L{{E}_{\varepsilon }}(G)=\sum\limits_{i=1}^{n}{|\mu _{i}^{'}-\frac{\zeta (G)}{n}|}.\]

\n Recall that $\zeta (G)/{n}$ is the average vertex eccentricity. In this paper, we fist calculate some basic properties and then establish some upper and lower bounds for $E{{L}_{\varepsilon }}(G)$.

\section{Main Results}

We know that, the ordinary and Laplacian graph eigenvalues obey the following relations:

$\sum\limits_{i=1}^{n}{{{\lambda }_{i}}}=0$; $\sum\limits_{i=1}^{n}{{{\lambda }_{i}}^{2}}=2m$

$\sum\limits_{i=1}^{n}{{{\mu }_{i}}}=2m$; $\sum\limits_{i=1}^{n}{{{\mu }_{i}}^{2}}=2m+\sum\limits_{i=1}^{n}{d{{({{v}_{i}})}^{2}}}.$

\n As the Laplacian spectrum is denoted by $\mu _{1}^{'},\mu _{2}^{'},...,\mu _{n}^{'}$, let $\nu _{i}^{'}=|\mu _{i}^{'}-\frac{\zeta (G)}{n}|$. Recall that the first Zagreb eccentricity index of a graph is denoted by ${{E}_{1}}(G)$ and is equal to the sum of square of all the vertices of the graph G. Thus, we have ${{E}_{1}}(G)=\sum\limits_{i=1}^{n}{\varepsilon {{({{v}_{i}})}^{2}}}$ (for details see \cite{xin11,luo14,de13a}). In the following, now we investigate some basic properties of $\mu _{i}^{'}$ and $\nu _{i}^{'}$.

\begin{lemma}
The eigenvalues $\mu _{1}^{'},\mu _{2}^{'},...,\mu _{n}^{'}$ satisfies the following relations

(i) $\sum\limits_{i=1}^{n}{\mu _{i}^{'}}=\zeta (G)$ and (ii) $\sum\limits_{i=1}^{n}{\mu {{_{i}^{'}}^{2}}}={E_1}(G)+2m$.
\end{lemma}
Proof. (i) Since, the trace of a square matrix is equal to the sum of its eigenvalues, we have

$\sum\limits_{i=1}^{n}{\mu _{i}^{'}}=tr(E{{L}_{\varepsilon }}(G))=\sum\limits_{i=1}^{n}{[\varepsilon ({{v}_{i}})-{{a}_{ii}}]}=\zeta (G).$

\n (ii) Again we have,
\begin{eqnarray*}
\sum\limits_{i=1}^{n}{\mu {{_{i}^{'}}^{2}}}&=&tr[(\varepsilon (G)-A(G)){{(\varepsilon (G)-A(G))}^{T}}]\\
           &=&tr[(\varepsilon (G)-A(G))(\varepsilon {{(G)}^{T}}-A{{(G)}^{T}})]\\
           &=&tr[\varepsilon (G)\varepsilon {{(G)}^{T}}-A(G)\varepsilon {{(G)}^{T}}-A{{(G)}^{T}}\varepsilon (G)+A(G)A{{(G)}^{T}}]\\
           &=&\sum\limits_{i=1}^{n}{\varepsilon {{({{v}_{i}})}^{2}}}+\sum\limits_{i=1}^{n}{{{\lambda }_{i}}^{2}}\\
          &=&{E_1}(G)+2m.
\end{eqnarray*}

\begin{lemma}
The eigenvalues $\nu _{1}^{'},\nu _{2}^{'},...,\nu _{n}^{'}$ satisfies the following relations

(i) $\sum\limits_{i=1}^{n}{\nu _{i}^{'}}=0$ and
(ii) $\sum\limits_{i=1}^{n}{\nu {{_{i}^{'}}^{2}}}={{E}_{1}}(G)-\frac{\zeta {{(G)}^{2}}}{n}+2m.$
\end{lemma}
\n\textit{Proof.}(i) We have from definition,
$\sum\limits_{i=1}^{n}{\nu _{i}^{'}}=\sum\limits_{i=1}^{n}{\left( \mu _{i}^{'}-\frac{\zeta (G)}{n} \right)}=\sum\limits_{i=1}^{n}{\mu _{i}^{'}}-\zeta (G)=0$. \qed

\n(ii) Again, similarly we have
\begin{eqnarray*}
\sum\limits_{i=1}^{n}{\nu {{_{i}^{'}}^{2}}}&=&\sum\limits_{i=1}^{n}{{{\left( \mu _{i}^{'}-\frac{\zeta (G)}{n} \right)}^{2}}}\\
         		&=&\sum\limits_{i=1}^{n}{\left[ \mu {{_{i}^{'}}^{2}}+\frac{\zeta {{(G)}^{2}}}{{{n}^{2}}}-2\mu _{i}^{'}\frac{\zeta (G)}{n} \right]}\\
   	             &=&{{E}_{1}}(G)+2m+\frac{\zeta {{(G)}^{2}}}{n}-2\frac{\zeta {{(G)}^{2}}}{n},
\end{eqnarray*}
which proves the desired result. \qed

\begin{lemma}
The eigenvalues $\nu _{1}^{'},\nu _{2}^{'},...,\nu _{n}^{'}$ satisfies the following relations
\[\left| \sum\limits_{i<j}{\nu _{i}^{'}\nu _{j}^{'}} \right|=\frac{1}{2}\left[ {{E}_{1}}(G)-\frac{\zeta {{(G)}^{2}}}{n}+2m \right].\]
\end{lemma}
\n\textit{Proof.} Since $\sum\limits_{i=1}^{n}{\nu _{i}^{'}}=0$, so we can write $\sum\limits_{i=1}^{n}{\nu {{_{i}^{'}}^{2}}}=-2\sum\limits_{i<j}{\nu _{i}^{'}\nu _{j}^{'}}$.
Thus, \[2\left| \sum\limits_{i<j}{\nu _{i}^{'}\nu _{j}^{'}} \right|=\sum\limits_{i=1}^{n}{\nu {{_{i}^{'}}^{2}}}={{E}_{1}}(G)-\frac{\zeta {{(G)}^{2}}}{n}+2m.\] Hence the desired result follows. \qed

As an example, in the following, we now calculate the eccentric version of Laplacian energy and corresponding spectrum of two particular type of graphs, namely, complete graph and complete bipartite graph.
\begin{ex}
Let $K_n$ be the complete graph with $n$ vertices, then
\[L_\varepsilon(K_n)=\left(
  \begin{array}{ccccc}
    1 & -1 & -1 & ... & -1 \\
    -1 & 1 & -1 & ... & -1 \\
    -1 & -1 & 1 & ... & -1 \\
    ... & ... & ... & ... & ... \\
    -1 & -1 & -1 & ... & 1 \\
  \end{array}
\right).\]
Its characteristic equation is
\[{{({{\mu }^{'}}-2)}^{(n-1)}}({{\mu }^{'}}-(2-n))=0.\]
Thus the eccentricity Laplacian spectrum of ${{K}_{n}}$ is given by

\[spe{{c}_{\varepsilon }}(G)=\left(
  \begin{array}{cc}
    (1-n) & 1 \\
    1 & (n-1) \\
  \end{array}
\right)\]
and hence $E{{L}_{\varepsilon }}(G)=2(n-1).$
\end{ex}

\begin{ex}
Let $K_{m,n}$ be the complete graph with $(m+n)$ vertices and $mn$ edges, then
\[L_\varepsilon(K_{n,n})=\left(
  \begin{array}{cccccccccc}
    2 & 0 & 0 & ... & 0 & -1 & -1 & -1 & ... & -1 \\
    0 & 2 & 0 & ... & 0 & -1 & -1 & -1 & ... & -1 \\
    0 & 0 & 2 & ... & 0 & -1 & -1 & -1 & ... & -1 \\
    .. & .. & .. & .. & .. & .. & .. & .. & .. & .. \\
    0 & 0 & 0 & ... & 2 & -1 & -1 & -1 & ... & -1 \\
    -1 & -1 & -1 & ... & -1 & 2 & 0 & 0 & ... & 0 \\
    -1 & -1 & -1 & ... & -1 & 0 & 2 & 0 & ... & 0 \\
    -1 & -1 & -1 & ... & -1 & 0 & 0 & 2 & ... & 0 \\
    .. & .. & .. & .. & .. & .. & .. & .. & .. & .. \\
    -1 & -1 & -1 & ... & -1 & 0 & 0 & 0 & ... & 2 \\
  \end{array}
\right).\]
Its characteristic equation is
\[{{({{\mu }^{'}}-2)}^{2(n-1)}}({{\mu }^{'}}-(2+n))({{\mu }^{'}}-(2-n))=0.\]
So, the eccentricity Laplacian spectrum of ${{K}_{n,n}}$ is given by

\[spe{{c}_{\varepsilon }}(G)=\left(
  \begin{array}{ccc}
    -n & n & 0 \\
    1 & 1 & 2(n-1) \\
  \end{array}
\right)\]
and hence $E{{L}_{\varepsilon }}(K_{n,n})=2n.$
\end{ex}

Note that, from the above two examples, the properties of the eigenvalues $\nu _{1}^{'},\nu _{2}^{'},...,\nu _{n}^{'}$ can be verified easily. In the following, next we investigate some upper and lower bounds of eccentricity version of Laplacian energy of a graph $G$.
\begin{Theorem}
Let G be a connected graph of order n and size m, then
\[E{{L}_{\varepsilon }}(G)\ge 2\sqrt{m+\frac{1}{2}({{E}_{1}}(G)-\frac{1}{n}\zeta {{(G)}^{2}})}.\]
\end{Theorem}

\n\textit{Proof.} We have from definition, $E{{L}_{\varepsilon }}(G)=\sum\limits_{i=1}^{n}{|\nu _{i}^{'}|}$. So we can write,
\[E{{L}_{\varepsilon }}{{(G)}^{2}}=\sum\limits_{i=1}^{n}{\nu {{_{i}^{'}}^{2}}+2\sum\limits_{i<j}{|\nu _{i}^{'}\nu _{j}^{'}|}}\ge \left( {{E}_{1}}(G)-\frac{\zeta {{(G)}^{2}}}{n}+2m \right)+2\sum\limits_{i<j}{|\nu _{i}^{'}\nu _{j}^{'}|}.\]
Hence using Lemma 3, the desired result follows. \qed

\begin{Theorem}
Let G be a connected graph of order n and size m; and $\nu _{1}^{'}$ and $\nu _{n}^{'}$ are maximum and minimum absolute values of $\nu _{i}^{'}$s, then
\[E{{L}_{\varepsilon }}(G)\ge \sqrt{n{{E}_{1}}(G)-\zeta {{(G)}^{2}}+2mn-\frac{{{n}^{2}}}{4}{{(\nu _{1}^{'}-\nu _{n}^{'})}^{2}}}.\]
\end{Theorem}

\n\textit{Proof.} Let ${{a}_{i}}$ and ${{b}_{i}}$, $1\le i\le n$ are non-negative real numbers, then using the Ozeki's inequality \cite{ozek68}, we have

\[\sum\limits_{i=1}^{n}{{{a}_{i}}^{2}}\sum\limits_{i=1}^{n}{{{b}_{i}}^{2}}-{{\left( \sum\limits_{i=1}^{n}{{{a}_{i}}{{b}_{i}}} \right)}^{2}}\le \frac{{{n}^{2}}}{4}{{({{M}_{1}}{{M}_{2}}-{{m}_{1}}{{m}_{2}})}^{2}}\]

\n where ${{M}_{1}}=\max ({{a}_{i}})$, ${{m}_{1}}=\min ({{a}_{i}})$, ${{M}_{2}}=\max ({{b}_{i}})$ and  ${{m}_{2}}=\min ({{b}_{i}})$.
Let ${{a}_{i}}=|\nu _{i}^{'}|$ and ${{b}_{i}}=1$, then from above we have
\[\sum\limits_{i=1}^{n}{|\nu _{i}^{'}{{|}^{2}}}\sum\limits_{i=1}^{n}{{{1}^{2}}}-{{\left( \sum\limits_{i=1}^{n}{|\nu _{i}^{'}|} \right)}^{2}}\le \frac{{{n}^{2}}}{4}{{(\nu _{1}^{'}-\nu _{n}^{'})}^{2}}.\]

Thus, we can write
\begin{eqnarray*}
E{{L}_{\varepsilon }}{{(G)}^{2}}&\ge& n\sum\limits_{i=1}^{n}{|\nu _{i}^{'}{{|}^{2}}}-\frac{{{n}^{2}}}{4}{{(\nu _{1}^{'}-\nu _{n}^{'})}^{2}}\\
 &=&n\left( {{E}_{1}}(G)-\frac{\zeta {{(G)}^{2}}}{n}+2m \right)-\frac{{{n}^{2}}}{4}{{(\nu _{1}^{'}-\nu _{n}^{'})}^{2}},
\end{eqnarray*}
from where the desired result follows.   \qed

\begin{Theorem}
Let G be a connected graph of order n and size m; and $\nu _{1}^{'}$ and $\nu _{n}^{'}$ are maximum and minimum absolute values of $\nu _{i}^{'}$s, then
\[E{{L}_{\varepsilon }}(G)\ge \frac{1}{(\nu _{1}^{'}+\nu _{n}^{'})}\left[ {{E}_{1}}(G)-\frac{\zeta {{(G)}^{2}}}{n}+2m-n\nu _{1}^{'}\nu _{n}^{'} \right].\]
\end{Theorem}

\n\textit{Proof.} Let ${{a}_{i}}$ and ${{b}_{i}}$, $1\le i\le n$ are non-negative real numbers, then from Diaz-Metcalf inequality \cite{diaz63}, we have
\[\sum\limits_{i=1}^{n}{{{b}_{i}}^{2}}+mM\sum\limits_{i=1}^{n}{{{a}_{i}}^{2}}\le (m+M)\left( \sum\limits_{i=1}^{n}{{{a}_{i}}{{b}_{i}}} \right).\]
where, $m{{a}_{i}}\le {{b}_{i}}\le M{{a}_{i}}$. Let ${{a}_{i}}=1$ and ${{b}_{i}}=|\nu _{i}^{'}|$, then from above we have
\[\sum\limits_{i=1}^{n}{|\nu _{i}^{'}{{|}^{2}}}+\nu _{1}^{'}\nu _{n}^{'}\sum\limits_{i=1}^{n}{{{1}^{2}}}\le (\nu _{1}^{'}+\nu _{n}^{'})\left( \sum\limits_{i=1}^{n}{|\nu _{i}^{'}|} \right).\]
Now, since \[E{{L}_{\varepsilon }}(G)=\sum\limits_{i=1}^{n}{|\nu _{i}^{'}|}\] and  $\sum\limits_{i=1}^{n}{\nu {{_{i}^{'}}^{2}}}={{E}_{1}}(G)-\frac{\zeta {{(G)}^{2}}}{n}+2m$, we get
\[{{E}_{1}}(G)-\frac{\zeta {{(G)}^{2}}}{n}+2m+n\nu _{1}^{'}\nu _{n}^{'}\le (\nu _{1}^{'}+\nu _{n}^{'})E{{L}_{\varepsilon }}(G),\]
from where the desired result follows.   \qed

\begin{Theorem}
Let G be a connected graph of order n and size m, then
\[E{{L}_{\varepsilon }}(G)\le \sqrt{\left[ n{{E}_{1}}(G)-\zeta {{(G)}^{2}}+2mn \right]}.\]
\end{Theorem}
\n\textit{Proof.} Using the Cauchy-Schwarz inequality to the vectors $(|\nu _{1}^{'}|,|\nu _{2}^{'}|,...,|\nu _{n}^{'}|)$ and $(1,1,...,1)$, we have
\[\sum\limits_{i=1}^{n}{|\nu _{i}^{'}|}\le \sqrt{n}\sqrt{\sum\limits_{i=1}^{n}{|\nu _{i}^{'}{{|}^{2}}}}\]
Thus from definition, we have
\begin{eqnarray*}
E{{L}_{\varepsilon }}(G)=\sum\limits_{i=1}^{n}{|\nu _{i}^{'}|}&\le& \sqrt{n}\sqrt{\sum\limits_{i=1}^{n}{|\nu _{i}^{'}{{|}^{2}}}}\\
   &=&\sqrt{n}\sqrt{\sum\limits_{i=1}^{n}{|\mu _{i}^{'}-\frac{\zeta (G)}{n}{{|}^{2}}}}=\sqrt{n}\sqrt{\left[ {{E}_{1}}(G)-\frac{\zeta {{(G)}^{2}}}{n}+2m \right]}.
\end{eqnarray*}
Hence the desired result follows.  \qed
\begin{Theorem}
Let G be a connected graph of order n and size m, then
\[E{{L}_{\varepsilon }}(G)\ge \frac{2\sqrt{\nu _{1}^{'}\nu _{n}^{'}}}{\nu _{1}^{'}+\nu _{n}^{'}}\sqrt{\left[ n{{E}_{1}}(G)-\zeta {{(G)}^{2}}+2mn \right]}.\]
\end{Theorem}

\n\textit{Proof.} We have, from Polya-Szego inequality \cite{poly72} for non-negatve real numbers ${{a}_{i}}$ and ${{b}_{i}}$, $1\le i\le n$
\[\sum\limits_{i=1}^{n}{{{a}_{i}}^{2}}\sum\limits_{i=1}^{n}{{{b}_{i}}^{2}}\le \frac{1}{4}{{\left( \sqrt{\frac{{{M}_{1}}{{M}_{2}}}{{{m}_{1}}{{m}_{2}}}}+\sqrt{\frac{{{m}_{1}}{{m}_{2}}}{{{M}_{1}}{{M}_{2}}}} \right)}^{2}}{{\left( \sum\limits_{i=1}^{n}{{{a}_{i}}{{b}_{i}}} \right)}^{2}}\]
where ${{M}_{1}}=\max ({{a}_{i}})$, ${{m}_{1}}=\min ({{a}_{i}})$, ${{M}_{2}}=\max ({{b}_{i}})$ and  ${{m}_{2}}=\min ({{b}_{i}})$. Let ${{a}_{i}}=|\nu _{i}^{'}|$ and ${{b}_{i}}=1$, then from above we have
\[\sum\limits_{i=1}^{n}{|\nu _{i}^{'}{{|}^{2}}}\sum\limits_{i=1}^{n}{{{1}^{2}}}\le \frac{1}{4}{{\left( \sqrt{\frac{\nu _{n}^{'}}{\nu _{1}^{'}}}+\sqrt{\frac{\nu _{1}^{'}}{\nu _{n}^{'}}} \right)}^{2}}{{\left( \sum\limits_{i=1}^{n}{|\nu _{i}^{'}|} \right)}^{2}}.\]
That is,
\begin{eqnarray*}
n\sum\limits_{i=1}^{n}{|\nu _{i}^{'}{{|}^{2}}}&\le& \frac{1}{4}{{\left( \sqrt{\frac{\nu _{n}^{'}}{\nu _{1}^{'}}}+\sqrt{\frac{\nu _{1}^{'}}{\nu _{n}^{'}}} \right)}^{2}}{{\left( E{{L}_{\varepsilon }}(G) \right)}^{2}}\\
        &=&\frac{1}{4}\frac{{{(\nu _{n}^{'}+\nu _{1}^{'})}^{2}}}{\nu _{n}^{'}\nu _{1}^{'}}{{\left( E{{L}_{\varepsilon }}(G) \right)}^{2}}.
\end{eqnarray*}
Since, $\sum\limits_{i=1}^{n}{\nu {{_{i}^{'}}^{2}}}={{E}_{1}}(G)-\frac{\zeta {{(G)}^{2}}}{n}+2m$, we get the desired result from above.     \qed

\section{Conclusion}

In this paper, we study different properties and bounds of eccentricity version of Laplacian energy of a graph $G$. It is found that, there is great analogy between the original Laplacian energy and eccentricity version of Laplacian energy, where as also have some distinct difference.

\section*{Competing Interests}

The author declares that there is no conflict of interests regarding the publication of this paper.

\end{document}